\documentclass[aps,twocolumn,preprintnumbers,nofootinbib,showpacs]{revtex4}
\usepackage{graphicx,amsmath}
\begin{document}
\title{TeV Colored Higgsinos in Alternative Grand Unified Theories}
\author{Kingman Cheung$^{1)}$ and Gi-Chol Cho$^{2)}$}
\affiliation{
$^1$
National Center for Theoretical Sciences, National Tsing Hua 
University, Hsinchu, Taiwan, R.O.C. \\
$^2$
Department of Physics, Ochanomizu University, Tokyo 112-8610, Japan}
\date{\today}

\begin{abstract}
Recently, a type of GUT models with an extra dimension
in AdS space can successfully solve the doublet-triplet problem, 
maintain proton stability, and allow the colored Higgs bosons and colored
Higgsinos in the TeV mass region.
We study the hadronic production and detection of these 
TeV colored Higgsinos and Higgs bosons.
If the colored Higgsino is lighter than the colored Higgs boson, 
the colored Higgs boson will decay into
a colored Higgsino and a gluino or a gravitino, and vice versa.  
The signatures would be stable massive charged particles with jets or
missing energies.
\end{abstract}
\preprint{NSC-NCTS-030608, OCHA-PP-207}
\maketitle

{\it Introduction.}
Grand unified theories (GUT) were introduced in 70's in an attempt to
unify the electromagnetic, weak, and strong interactions into a single
theory.  The most well-known model is based on the SU(5) symmetry 
\cite{georgi}, which is the smallest single gauge group that can incorporate
the SU(3) color, SU(2) weak, and the U(1) hypercharge
interactions.  In addition, the matter fermions are also grouped into
representations of the SU(5), and thus achieving the quark-lepton
unification to some extent.  One of the most successful features of GUT is
the electric charge quantization.  
The SU(5) symmetry is broken at some high scales to 
SU(3)$_C\times$SU(2)$_L\times$U(1)$_Y$, giving rise to heavy $X,Y$ gauge
bosons.  The exchanges of the $X,Y$ gauge bosons 
induce proton decay (e.g. $p\to e^+ \pi^0$) \cite{weinberg}
via dimension-six operators $(QQ)(QL)$, which are suppressed by two powers of
the GUT breaking scale: $g^2_{X,Y}/M^2_{GUT}$.  
Other dimension-six proton decay operators can also arise from 
exchanges of the Higgs color triplets.
In the supersymmetric version of GUT \cite{susy-gut},  
there are additional sources of 
proton decay.  Exchanges of the colored Higgsinos, the supersymmetric 
partners of the Higgs color triplet, give rise to dimension-five
operators $(QQ)(\widetilde{Q}\widetilde{L})$, which are only suppressed 
by one power of the GUT breaking scale \cite{five}.

Unobserved proton decay has pushed the mass of the
colored Higgs bosons and Higgsinos to at least a few $\times 10^{16}$
GeV \cite{hitoshi}.  Compared to the mass of the weak doublet Higgs fields,
which is fixed at around $O(100)$ GeV, the mass of the color triplet is 
$14-15$ orders of magnitude heavier.  Such a large mass difference between
the triplet and the doublet is a serious
problem in supersymmetric GUT (SUSY-GUT), 
dubbed the doublet-triplet splitting problem \cite{susy-gut}. 
This problem arises because the weak-doublet Higgs fields,
 which are responsible for the electroweak symmetry breaking,
and the color triplet belong to the same $\textbf{5}$  
and $\bar{\textbf{5}}$ representations of SU(5):
\begin{eqnarray}
H(\textbf{5}) &=& (H_C, H_u),  \nonumber \\
H(\bar{\textbf{5}}) &=& (\bar{H}_C, H_d) \nonumber \,
\end{eqnarray}
where the weak doublets $H_u$ and $H_d$ are responsible for 
the up- and the down-type quark (lepton) masses, respectively. 
 Therefore, the 
color triplet and the weak doublet will naturally 
have the same Yukawa couplings to matter fermions.  The color-triplet
fields will cause the proton decay.  That is the reason why the color
triplet has to be very heavy (a few $\times 10^{16}$ GeV) to avoid the
proton decay constraint.  The doublet-triplet problem is perhaps the most
undesirable problem of the SUSY-GUT.
Most attempts to the doublet-triplet problem in literature
have been focused on how to naturally explain the 
hierarchy between the weak-doublet and the color-triplet
masses after the GUT symmetry breaking~\cite{DT-splitting}. 

An alternative approach to the doublet-triplet 
splitting problem
is to suppress the Yukawa couplings of the color-triplet Higgs fields 
to matter fermions in order to preserve the proton longevity.
Thus, no mass splitting between the weak-doublet and color-triplet 
Higgs fields is required.  The color-triplet mass 
can be as low as $O(TeV)$, and 
can be copiously produced at the upcoming LHC \cite{CC}.
The suppression of Yukawa couplings of the triplet can be achieved by
some clever group structures \cite{Dvali:1995hp} or by extra dimension 
setups \cite{Haba:2002if,nomura}

A particularly interesting mechanism is by orbifolding 
 \cite{kawamura}. When one compactifies the extra 
dimensions, one can assign special boundary conditions to various 
components of a multiplet.  After compactification, some components are
automatically zero at the orbifold fixed point while other components are
not, thus breaking the symmetry and naturally splitting the multiplet.  
If the multiplet consists of the Higgs color triplet and weak doublet,
one can make the color triplet automatically zero on the fixed point,
where matter fermions reside, to achieve proton stability and natural
doublet-triplet splitting.
In the model by Goldberger et al. \cite{nomura} (see also earlier works
\cite{pomarol}), they started from
the Randall-Sundrum scenario \cite{randall}: a slice of AdS space
with two branes (the Planck brane and the TeV brane), one at each end.
%The extra dimension is compactified on an $S^1/Z_2$ orbifold.  
The hierarchy of scales is generated by the AdS warp factor $k$, which is
of order of the five-dimensional Planck scale $M_5$.
%such that the 4D Planck scale is given by $M^2_{\rm Pl} \sim M_5^3/k$.  
The fundamental
scale on the Planck brane is $M_{\rm Pl}$ while the fundamental scale
on the TeV brane is rescaled to TeV by the warp factor: $T\equiv k
e^{-\pi k R}$, where $R$ is the size of the extra dimension.
The model is a 5D supersymmetric SU(5) gauge theory compactified on the 
orbifold $S^1/Z_2$ in the AdS
space.  The boundary conditions break the SU(5) symmetry and provide
a natural mechanism for the Higgs doublet-triplet splitting.
The Planck brane respects the SM gauge
symmetry while the TeV brane respects the SU(5) symmetry.  
By the boundary conditions the
wave-function of the color-triplet Higgs fields is zero
at the Planck brane, on which the matter fermions reside, while the
doublet Higgs fields are nonzero at the Planck brane and give Yukawa
couplings to the matter fermions.  Thus, the excessive proton decay
via the color-triplet Higgs fields is highly suppressed.
The mass of the color-triplet fields is given by the warp factor to be in 
TeV scale.

In this Letter, we calculate the production and describe the detection of 
the TeV colored Higgs bosons and Higgsinos in hadron colliders. 
The colored Higgs bosons and Higgsinos will
give rise to a novel signature like ``heavy muons''.
Depending on their masses, the colored Higgs boson will decay into the
colored Higgsino and a gluino or a gravitino (in this kind of models the
gravitino is very light \cite{nomura,pomarol}), and vice versa.  
The present work has 
non-trivial improvements over our previous work \cite{CC}: 
(i) we take into account production
of both colored Higgs bosons and Higgsinos, (ii) the signature
of colored Higgs boson decaying to colored Higgsino and gluino or gravitino
gives rise to jets or missing energies, and (iii) increased event rates
improve significantly the sensitivity at the LHC.  

{\it Production of colored Higgsinos.}
Let us denote the SUSY partner of $H_C$ by $\widetilde{h}_C$ and
that of $\bar H_C$ by $\widetilde{\bar h}_C$
(anti-particles are denoted by $H_C^*$ and $\overline{\widetilde{h}_C}$,
and $\bar H_C^*$ and $\overline{\widetilde{\bar h}_C}$, respectively.)
As we already mentioned, the colored Higgs bosons and Higgsinos
do not have sizable 
Yukawa couplings to the SM fermions in order to suppress the fast 
proton decay. 
Thus, the only allowed production channels of the colored Higgs bosons
and Higgsinos in hadronic collisions are via the SU(3)$_C$ invariant 
interactions
\begin{eqnarray}
{\cal L} &=& -i g_s H_C^* {\stackrel{\leftrightarrow}{\partial}}_\mu 
H_C T^a A^{a\mu} 
     + g_s^2 T^a T^b H_C^* H_C A^a_\mu A^{b\mu} \nonumber \\
&& - g_s\, T^a A^a_\mu \,
       \overline{\widetilde{h}_C} \gamma^\mu   \widetilde{h}_C \, 
    \nonumber \\
&& -\sqrt{2} g_s \left(
H_C^* \, \overline{\widetilde{g}^a} \,T^a \, \widetilde{h}_C + 
\overline{\widetilde{h}_C} \,
 T^a \,\widetilde{g}^a \,  H_C \right),
\end{eqnarray} 
%%%-------------
where $T^a$ is the generator of SU(3), and 
$A {\stackrel{\leftrightarrow}{\partial}}_\mu B 
\equiv A(\partial_\mu B) - (\partial_\mu A)B$. 
The last line is the matter-gaugino interaction with the gluino.
The interactions for $\bar{H}_C, \widetilde{\bar{h}}_C$ are the same as 
$H_C, \widetilde{h}_C$, respectively.
It is understood that $\widetilde{h}_C$ is a
left-handed field.
The production of the colored Higgs bosons and Higgsinos 
in the lowest order is via the $s$-channel $q\bar{q}$ annihilation 
and the glue-glue fusion.   The production of the colored Higgs bosons
has been calculated in Ref. \cite{CC}.  Here we present the formulas
for the production of the colored Higgsinos:
\begin{eqnarray}
\frac{d\hat \sigma}{d \cos\theta^*} 
\left(q \bar q \to \widetilde{h}_C \overline{\widetilde{h}_C} \right ) &=& 
\frac{\pi \alpha_s^2}{ 9 \hat s} \, \beta \, \left( 1 - 
 \frac{2 \hat t_1 \hat u_1}{\hat s^2} \right ) \\
\frac{d\hat \sigma}{d \cos\theta^*} 
\left(g g  \to \widetilde{h}_C \overline{\widetilde{h}_C} \right ) &=& 
\frac{\pi \alpha_s^2}{ 96 \hat s} \, \beta \, \Biggr [
 \left( \frac{4 \hat s^2}{ \hat t_1 \hat u_1} - 9 \right )\; \nonumber  \\ 
&& \hspace{-1.2in}
\left.
\times \left ( \frac{\hat s^2} {\hat t_1 \hat u_1}-  2 \right )\,
       \left ( \frac{\hat t_1 \hat u_1}{\hat s^2} 
         -  \frac{ m^2_{\widetilde{h}_C} }{\hat s} \right )
   + \frac{ 7 m^2_{\widetilde{h}_C}}{ \hat s }
 \right ]
\end{eqnarray}
where $\beta = \sqrt{ 1 - 4 m^2_{\widetilde{h}_C}/ \hat s}$, 
$\hat t_1 = \hat t - m^2_{\widetilde{h}_C}$,
$\hat u_1 = \hat u - m^2_{\widetilde{h}_C}$, and
$\theta^*$ is the scattering angle in the parton rest frame and is \
related to $\hat t_1$ by $\hat t_1 = -\frac{\hat s}{2}
 (1-\beta \cos\theta^*)$. 
Note that the expressions for the production cross section of
the $\widetilde{\bar h}_C \overline{\widetilde{\bar h}_C}$ 
pair are the same as 
the $\widetilde{h}_C \overline{\widetilde{h}_C}$ pair.
%If the mass of $\widetilde{\bar h}_C $ is the same as $\widetilde{h}_C$, 
%the sum of the cross sections would be doubled.  
In the minimal SUSY SU(5), they have exactly the same mass. 
Even beyond the minimal model, since there is no particular reason why
their masses should be very different, we simply take them to be equal 
and the results present in the following take into account both 
$\widetilde{h}_C$ and $\widetilde{\bar h}_C$.

\begin{figure}[th!]
\includegraphics[width=3.4in]{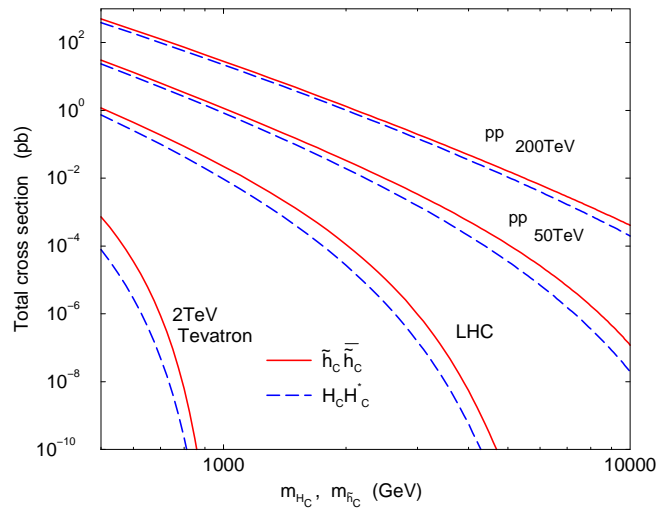}
\caption{\small\label{total}
Total cross sections for the production of the colored Higgsino and 
Higgs boson pair at 
the Tevatron, LHC, and $pp$ collisions at 50 and 200 TeV.
}
\end{figure}

In the calculation, we employ the parton distribution function of CTEQ v.5
(set L) \cite{Lai:1999wy} and the 
one-loop renormalized running strong coupling
constant with $\alpha_s (M_Z) =0.119$.  
%The total cross section at a center-of-mass energy $\sqrt{s}$ 
%is obtained by convoluting the partonic cross 
%sections in Eqs. (\ref{gg}) and (\ref{qq}) with the parton 
%distribution functions:
%\begin{equation}
%\sigma(s) = \int_{4m_{H_C}^2/s}^{1}\; d \tau \;\int_{\tau}^{1} \;\frac{dx}{x}\;
%f(\tau/x) \,f(x) \; \hat \sigma( \hat s) \;,
%\end{equation}
%where $\hat s = \tau s$ is square of the 
%center-of-mass energy of the parton-parton scattering.
%%
We show in Fig. \ref{total} the production cross sections of colored Higgsinos
and Higgs bosons vs the masses $m_{\widetilde{h}_C}$ and $m_{H_C}$
at the Tevatron ($p\bar p$ collisions), at 
the LHC ($pp$ collisions), and in $pp$ collisions at $\sqrt{s}=50,200$ TeV
(the lower and upper energy range of the VLHC \cite{vlhc}.)

%Is there any indirect constraints on the colored Higgs bosons from 
%the high-energy experiments? 
%Let us consider the $Z$-pole experiments at LEP1 and SLC 
%where the radiative corrections to the gauge boson propagators 
%and the $Z \to f\bar{f}$ ($f$ denotes quarks or leptons) 
%vertices are severely constrained. 
%The contributions to the gauge boson propagators are summarized 
%by the $S,T,U$ parameters~\cite{Peskin}. 
%It is well known that the SU(2)$_L$ singlet scalars do not contribute 
%to the $S,T,U$ parameters~\cite{Cho:1999km} such that the colored Higgs 
%boson is free from the constraints.
%Since the Yukawa interactions of $H_C$ and $\bar{H}_C$ to the quarks and 
%leptons are highly suppressed, they do not contribute to the 
%$Z\to f\bar{f}$ processes.
%By the same reason, there is no constraint on the colored Higgs 
%bosons from flavor physics experiments. 
%We, therefore, conclude that no indirect constraints are implied
%for the colored Higgs mass from current experiments. 

%%% Decay %%%%%%%%%%%%
{\it Detection of colored Higgsinos.}
Depending on their masses, the colored Higgs boson will decay into the
colored Higgsino and a gluino or a gravitino, and vice versa. 
We have the following scenarios
\begin{eqnarray}
(i) m_{H_C} > m_{\widetilde{h}_C} & & 
         H_C \to \widetilde{h}_C  \; +\;
     \widetilde{g} /\widetilde{G} \nonumber \\
(ii) m_{\widetilde{h}_C} > m_{H_C} && 
         \widetilde{h}_C \to H_C  \;+\;
 \widetilde{g} /\widetilde{G} \nonumber\;.
\end{eqnarray}
The 95\% C.L. limit on the mass of the gluino is 
195 GeV independent of the squark mass \cite{cdf-gl}.  We use
$m_{\widetilde{g}} = 220$ GeV.  In the case (i), if
 $m_{H_C} > m_{\widetilde{h}_C} + m_{\widetilde{g}}$, then the colored 
Higgs boson will decay into the colored Higgsino and the gluino, 
otherwise it will decay into the colored Higgsino and a gravitino.  
(Vice versa for the case (ii) $m_{\widetilde{h}_C} > m_{H_C} + 
m_{\widetilde{g}}$.)

In either case, the lighter of the colored Higgs boson and Higgsino
will hadronize into a stable massive particle,
which is electrically either neutral or charged.  In the following
discussion, let the colored Higgsino be the lighter one
(the reverse is similar.)
Both the neutral and charged states will undergo hadronic energy loss 
in the detector.
Although it is strongly interacting, hadronic energy loss is
negligible because of the small momentum transfer
between the massive particle (TeV) and the nucleon.  
Thus, the energy loss via hadronic collisions does not lead to detection
of the massive particle.  
On the other hand, the charged state will also undergo
ionization energy loss $dE/dx$.
In Ref. \cite{CC}, we have shown that $dE/dx$ as a function of $\beta\gamma$
almost has no explicit dependence on the mass of the penetrating particle, 
especially in the range $ 0.1 < \beta \gamma < 1$.
Therefore, when $dE/dx$ is measured in an experiment, the $\beta\gamma$ can
be deduced, which then gives the mass of the particle if the momentum $p$ is
simultaneously 
measured.  Hence, $dE/dx$ is a good tool for particle identification
for stable massive charged particles (MCP). 

In fact, the CDF Collaboration did a few searches for stable MCPs
\cite{cdf1}. 
The CDF analyses require 
that the particle produces a track in the central tracking system
and penetrates to the outer muon chamber.
(In the Run II analysis, the requirement of ionization in the outer muon
chamber may not be necessary.)
The CDF requirement on $\beta\gamma$ is
\begin{equation}
\label{cut}
0.25 \;\; \alt \;\;\; \beta \gamma \;\;\; < \;\; 0.85  \;.
\end{equation}
The lower limit is to make sure that the penetrating particle can make it to 
the outer muon chamber while the upper limit makes sure that the ionization
loss in the tracking system is sufficient for a detection.   We shall 
employ the same kinematical cut.  We have verified in Ref. \cite{CC}
that such a cut on $\beta\gamma$ is also valid for a 1 TeV particle.
Since the lower cut on $\beta\gamma=p/M$ is $0.25$, the momentum cut is
$250$ GeV for a 1 TeV particle.  Such a cut on momentum already
makes it background free from $\mu^\pm, \pi^\pm, K^\pm$.
Another configuration cut due to the detector (both CMS and Atlas) is 
\begin{equation}
\label{eta}
|\eta| < 2.5 \;.
\end{equation}
We also assume an efficiency of 80\% for each stable MCP
to be detected by the central
tracking system and the outer muon system.
This efficiency is in addition to the cuts on $\eta$ and $\beta\gamma$.

We assume that the gluino decays into quarks and neutralino in the usual 
neutralino-LSP scenario.  Thus, if gluinos appear in the decay of colored
Higgs bosons, there will also be jets and missing 
energies in the final state.  Therefore, one or two MCPs 
with or without jets can be seen in the final state.  To definitely see a
jet we impose
\begin{equation}
p_{Tj} > 50\; {\rm GeV}\;, \;\; |\eta_j| < 2.5\;.
\end{equation}
On the other hand, if the colored Higgs boson is not heavy enough to decay 
into the colored Higgsino and the gluino, it will decay into the 
colored Higgsino and the gravitino.  In this type of models, the gravitino
is often sub-eV \cite{nomura,pomarol}. 
In this case, there will be one or two MCPs with or without
missing energies.  However, since the MCP may not be fully ionized in the
detector, it is rather difficult to determine
the missing energy caused by the gravitino.  

We estimate the 
number of observed events in the production of colored Higgsinos and
Higgs bosons.  Throughout, we use $m_{\widetilde{g}}=220$ GeV,
$m_{\widetilde{\chi}^0_1}=120$ GeV,
and $m_{H_C} = m_{\widetilde{h}_C} + 100,\,300,\,500$ GeV for the 
case when the colored Higgs boson is heavier, and vice versa.  
We also employ the following factors in estimating the event rates:
\begin{itemize}
\item[(i)]
a probability of $1/2$ that the colored particle will hadronize into 
an electrically charged particle,
\item[(ii)]
an efficiency factor of $0.8$ for each detected track, and
\item[(iii)]
both channels of $H_C H_C^*$ and $\bar H_C \bar H_C^*$, and
$\widetilde{h}_C$ pair and $\widetilde{\bar h}_C$ pair production 
are added.
\end{itemize}
Since the search is background free, the discovery or evidence of existence
for the colored Higgs bosons and Higgsinos 
depends crucially on the number of observed
events, which we choose to be 10--20 events.  
We show the event rates for the LHC in Table \ref{table1} for both
cases: $m_{H_C} > m_{\widetilde{h}_C}$ and 
$m_{H_C} < m_{\widetilde{h}_C}$.

{\it Conclusions.}
The presence of light color-triplet Higgs fields in TeV mass 
scale is a novel feature for the alternative kind of GUT, instead of 
proton decay.  This is made possible through some mechanisms to suppress 
the Yukawa couplings of the triplets to matter fermions; in particular
the orbifolding in an AdS space naturally suppresses proton decay and gives
TeV color triplet.
The striking signature of these TeV colored Higgs bosons and Higgsinos
would be stable
massive charged particles, ``heavy muons'', producing a 
track in the central tracking system and penetrating to the outer muon 
system.  Such a signature is background free and gives a clean indication
of massive charged particles (MCP).
We have calculated the event rates for 
various final states (1 MCP, 2MCP, 1MCP$+$jets, and 2MCP$+$jets) at 
the LHC.  The LHC with an
accumulated luminosity of 100 fb$^{-1}$ is sensitive to almost 2 TeV colored
Higgsinos and  Higgs bosons.

%It is well known that the success of gauge coupling unification 
%in the minimal supersymmetric standard model (MSSM) could be preserved 
%if complete multiplets of SU(5) (e.g.,  $\textbf{5}$ or $\textbf{10} 
%\cdots$) are added to the spectrum of MSSM. 
%When $H_C$ and $\bar{H}_C$ are in TeV scale, there must be a vector pair 
%of weak doublets in the same scale to form
%the $\textbf{5}$ and $\bar{\textbf{5}}$ multiplets so that the gauge 
%coupling unification is unaltered. 

We also note that the colored Higgs bosons and Higgsinos are 
SU(2)$_L$ singlets and thus do not contribute 
to the $S,T,U$ parameters~\cite{Cho:1999km}.
In addition, since their Yukawa couplings to quarks and leptons
are highly suppressed, they do not contribute to the 
$Z\to f\bar{f}$ processes or any other flavor processes. 
Therefore, there are no existing constraints on these particles, except 
for some direct search limits on stable massive charged particles \cite{cdf1}.
Observation of such color-triplet fields of TeV mass is definitely a signal
for the alternative GUT.

The work of G.C.C. is supported in part by the Grant-in-Aid for 
Science Research, Ministry of Education, Science and Culture, 
Japan (No.15740146). K.C. is supported the National Center for Theoretical
Sciences under a grant from the National Science Council of Taiwan R.O.C.

%%%%%%%%%% Table %%%%%%%%%%%%%%
\begin{table*}[!tbp]
\caption{\label{table1} \small
Event rates of massive charged particles (MCP) due to
pair production of colored Higgs bosons $H_C H_C^*$ and
$\bar H_C \bar H_C^*$ and colored Higgsinos $\widetilde{h}_C 
\overline{\widetilde{h}_C}$ and $\widetilde{\bar h}_C 
\overline{\widetilde{\bar h}_C}$ at the LHC with an integrated
luminosity of 100 fb$^{-1}$.
}
\medskip
\begin{ruledtabular}
\begin{tabular}{cccccc}
$m_{H_C}$ (TeV) & $m_{\widetilde{h}_C}$ (TeV) & 1 MCP & 2 MCP & 1 MCP $+$jets
                                                             & 2 MCP $+$jets \\
\hline
\multicolumn{6}{c}{$m_{H_C} > m_{\widetilde{h}_C}$}\\
$1.1$ & $1.0$ & $641$  &  $99$ & - &-\\
$1.3$ & $1.0$ & $557$  &  $85$ & $34$ & $5$ \\
$1.5$ & $1.0$ & $532$  &  $80$ & $11$ & $1$ \\
\hline
$1.3$ & $1.2$ & $208$  &  $35$ &- &-\\
$1.5$ & $1.2$ & $185$  &  $30$ & $11$ &$2$ \\
$1.7$ & $1.2$ & $177$  &  $29$ & $4$  &$1$ \\
\hline
$1.5$ & $1.4$ & $74$  &  $13$ &- &-\\
$1.7$ & $1.4$ & $66$  &  $12$ & $4$ & $1$ \\
$1.9$ & $1.4$ & $64$  &  $11$ & $1$ & $0$ \\
\hline
$1.7$ & $1.6$ & $28$  &  $5$ &- &-\\
$1.9$ & $1.6$ & $25$  &  $5$ & $1$ & $0$ \\
$2.1$ & $1.6$ & $24$  &  $5$ & $1$ & $0$ \\
\hline

\hline
\multicolumn{6}{c}{$ m_{\widetilde{h}_C} > m_{H_C}$}\\
$1.0$ & $1.1$ & $525$  & $83$  & -&-\\
$1.0$ & $1.3$ & $330$  & $54$   & $100$ & $14$ \\
$1.0$ & $1.5$ & $262$  & $44$   & $36$  & $4$  \\
\hline
$1.2$ & $1.3$ & $167$  & $29$   &  -     & -\\
$1.2$ & $1.5$ & $102$  & $18$   & $37$ & $6$ \\
$1.2$ & $1.7$ & $78$   & $14$   & $14$  & $2$ \\
\hline
$1.4$ & $1.5$ & $58$   & $11$   &  -    & -\\
$1.4$ & $1.7$ & $35$   & $7$    &  $14$ & $3$ \\
$1.4$ & $1.9$ & $26$   & $5$    &  $6$ & $1$ \\
\hline
$1.6$ & $1.7$ & $22$   & $4$   &  -    & -\\
$1.6$ & $1.9$ & $13$   & $3$    &  $6$ & $1$ \\
$1.6$ & $2.1$ & $9$   & $2$    &  $2$ & $0$
\end{tabular}
\end{ruledtabular}
\end{table*}

%%%%%%%%%%%%%%%%%%%%%%%%%%%%%%%%%%%%%%%%%%%%%%%%%%%%%%%%%%%%%%
%%%%%%%         BIBLIOGRAPHY           %%%%%%%%%%%%%%%%%%%%%%%
%%%%%%%%%%%%%%%%%%%%%%%%%%%%%%%%%%%%%%%%%%%%%%%%%%%%%%%%%%%%%% 

\end{document}